\documentclass[showpacs,prb]{revtex4}
\usepackage{graphicx}
\begin{document}

\title{Structural and electronic properties of MgO nanotube clusters}
\author{G. Bilalbegovi{\'c}}
\affiliation{Department of Physics, University of Rijeka, Omladinska 14, 51000 Rijeka, Croatia}

\date{to be published in Phys. Rev. B}

\begin{abstract}
Finite magnesium oxide nanotubes are investigated.  Stacks of
four parallel squares, hexagons, octagons, and decagons are constructed and studied by the pseudopotential density functional theory within the local-density approximation.
Optimized structures are slightly distorted stacks of polygons.
These clusters are insulators and
the band gap of $8.5$ eV is constant over an investigated range of the diameters of stacked polygonal rings.
Using the L{\" o}wdin population analysis
a charge transfer towards the oxygen atoms is estimated as $1.4$,
which indicates that the mixed ionocovalent bonding exists in
investigated MgO nanotubes.
\end{abstract}

\pacs{61.46.+w, 73.22.-f; 36.40.Cg, 81.07.De}

\maketitle

\section{Introduction}
\label{sec:intro}

Studies of nanotubes and nanowires are important for recent research in nanotechnology.
Experimental and theoretical investigations of cylindrical nanostructures for
many different organic and inorganic materials already exist. \cite{Avouris}
As new arrangements of atoms
these structures are interesting for basic science.
There is also a possibility to use cylindrical nanostructures as electrical and
mechanical parts in nanodevices. Some technological applications of nanotubes and nanowires already exist,
most notably for carbon nanotubes.
Magnesium oxide is important in applications. It is known as an inert
material with a high melting point, as a catalyst in important chemical reactions, and as a good substrate for many chemicals, for example metals, group
III-V elements, and high-temperature superconductors.
MgO is also used as an optical material and a component of composites.
A bulk magnesium oxide crystallizes in the rocksalt structure.
It is an insulating ionic simple oxide.

Properties of materials at a finite scale of lengths are often different in comparison to their bulk.
For example, a pronounced covalent contribution to the ionic bonding
exists in small MgO nanoparticles, whereas almost pure ionic bonds are typical for the bulk of this compound.
Therefore, a crossover from a mixed covalent and ionic bonding to dominantly ionic one occurs in MgO nanostructures of an intermediate size.
Small clusters of MgO have been investigated by several experimental \cite{Saunders,PJZiemann}
and theoretical methods.
\cite{PJZiemann,Kohler,Moukouri,Noguera,Malliavin,Roberts,Recio,Puente,Wilson,Veliah,FCalvo}
A study of MgO clusters and their nucleation from the gas phase has been used to
explore one possible process of a
cosmic iron-magnesium-silicate dust formation in the last phase of life for some stars.
\cite{Kohler}
Mass spectroscopy and collision-induced-fragmentation  measurements
have been performed on sputtered MgO cluster ions.
\cite{Saunders}
Results have been shown that for
small (MgO)$_n$ clusters especially pronounced peaks exist at intervals of $n=3$
(i.e., for $n=6,9,12,$ and $15$). These peaks indicate the most stable structures.
Laser-ionization time-of-flight mass spectrometry measurements on (MgO)$_n$ clusters produced in
a gas aggregation source have also been carried out.
\cite{PJZiemann}
The clusters of enhanced stabilities were found for $n=2,4,6,9,12,$ and $15$.
Ziemann and Castleman have been performed calculations using two polarizable ion shell models.
\cite{PJZiemann}
Their   results
have been shown that the  most stable structures  are compact cubelike  ones  in
the  model with ion charges  $\pm 1$. Conversely, spherical geometries  composed of
hexagons and squares have been found in the second model with ion charges $\pm 2$.
Hollow and compact cubelike clusters have been also found in various other calculations.
\cite{Kohler,Moukouri,Noguera,Malliavin,Roberts,Recio,Puente,Wilson,Veliah,FCalvo}
Cubelike structures are similar to pieces of the bulk MgO face-centered cubic lattice.
These cubelike clusters prevail when the number of particles increases.
Planar geometries, as well as spherical and cylindrical three-dimensional hollow ones, have also been found.
A  selfconsistent   tight     binding   calculations    of    structural    and
electronic properties of (MgO)$_n$   clusters   have been  carried out  by  Moukouri
and Noguera. \cite{Moukouri,Noguera}
They  have  been  calculated  that  planar  polygonal  rings
are stable configurations up to $n=6$.
The (MgO)$_n$ clusters of the same size have been investigated by the density functional DMOL package.
\cite{Malliavin}
These  results have  been
shown that investigated small (MgO)$_n$ clusters are the  most stable for $n=3$  and $n=6$.
Structures of MgO clusters have been studied by a genetic algorithm optimization
method within a simple empirical rigid ion model.
\cite{Roberts}
The variation of cluster geometry with the charges $\pm q$ on the ions have been investigated. When the charge changes from $q=1$ to $q=2$, the structure evolves from compact cubelike to hollow pseudo-spherical one.
Stacks of polygonal (MgO)$_n$ rings have been found  by the Hartree-Fock calculations for small clusters,
\cite{Recio,Puente} and by optimization using the Born-Mayer potential. \cite{PJZiemann}
Using an empirical compressible-ion interaction model potential and coordination number-dependent oxide polarizabilities Wilson  has been found stable (MgO)$_n$ nanotubes for $n < 30$. \cite{Wilson}
Electronic and related structural properties of these MgO  cylindrical hollow structures, i.e., nanotube clusters, have not been investigated.

Several methods of fabricating MgO nanowires have been developed. Boron oxide assisted catalytic growth has been used to synthesize MgO nanowires  with uniform diameter distribution of about $20$ nm, and a typical length of several tens of
micrometers.
\cite{Tang}
Sometimes, nanowires with length of several hundred micrometers have been observed.
X-ray diffraction, scanning electron microscopy, and high-resolution transmission electron microscopy techniques
have been revealed that these nanowires are single-crystal fcc structures.
MgO nanowires with diameters of $(15-20)$ nm, lengths of up to several hundred micrometers, and the fcc structure
have been also prepared using a vapor-phase precursor method starting from the MgB$_2$ powders.
\cite{Yin}
Cubic MgO nanowires with diameters of about 100 nm and lengths of up to several micrometers have been synthesized by heating the Mg powder above its melting point in a flow of 20\% O$_2$ and 80\% Ar.
\cite{Dang}
Recently MgO nanotubes have been synthesized using a liquid metal assisted growth.
\cite{YBLi}
They have been filled with gallium and
this nanoscopic device is proposed as a thermometer with better properties than previously constructed one based on a carbon nanotube. \cite{YGao}
The fabricated MgO nanotubes are cubic, several micrometers long and exhibit both interior and exterior square cross sections.
Their outer diameters are $(30-100)$ nm, whereas interior ones are $(20-60)$ nm.
MgO nanorods have been used to produce columnar defects in high-temperature superconductors.
\cite{PYang,CMLieber}
They could pin magnetic flux lines and prevent thermally activated flux flow.
Small MgO nanotubes studied in this work are also candidates for applications as components in the bulk, film, or wire composites of superconducting and other materials.

In this work MgO nanotubes having between $16$ and $40$ atoms, and made of stacks of four polygons are investigated.
By stacking four polygons on top of each other and optimizing these structures, slightly distorted nanotubes
of the length from $0.5782$ nm to $0.5925$ nm are formed.
The structure, energetics, and electronic properties of these MgO nanotube clusters are studied using computational methods based on a pseudopotential density functional theory in the local density approximation (LDA).
The rest of the paper is organized as follows. Section \ref{sec:2}  gives
an outline of the method. In Sec.~\ref{sec:3}  the results and discussion are presented.
Conclusions are summarized in Sec.~\ref{sec:4}.

\section{Computational Method}
\label{sec:2}

In this work  \textit{ab initio} density functional theory calculations
\cite{Kohn,Sham}
are performed to study the structural and electronic
properties of finite magnesium oxide nanotubes.
The plane-wave pseudopotential method is applied. \cite{SdeGironcoli,SBaroni}
The density functional theory within the same plane-wave pseudopotential framework  has already been used to calculate properties of the bulk MgO under high-temperature and high-pressure conditions which are important for
geophysics and planetary physics.
\cite{BBKarki,Wentzcovitch}
Recent density functional theory study of the high-pressure
behavior of MgO at the generalized gradient approximation level (GGA) \cite{Oganov}
has shown an agreement with  the LDA calculations of Refs.
\cite{BBKarki,Wentzcovitch}
Within the same plane-wave pseudopotential model as used in this work,
it has been shown that the LDA approximation for the bulk of MgO and other alkaline-earth oxides yields good agreement with available structural and vibrational experimental data.\cite{OSchutt}

In this calculation LDA and parametrization of Perdew and Zunger for the exchange-correlation energy is applied. \cite{Perdew}
The pseudopotentials generated by the method of von Barth and Car for magnesium \cite{Car,Wentzcovitch},
and by one of Troullier and Martins for oxygen \cite{Martins} are used.
The configurations $2s^2$$2p^4$ of O, and $3s^2$ of Mg are taken as valence states.
The calculations are performed with a kinetic-energy cutoff of $60$ Ry.
In the studies of MgO under geophysical conditions  \cite{Wentzcovitch,BBKarki} the cutoff of $90$ Ry  has been used, and
this is important for a behavior of materials at high pressures.
In the studies presented in this work (as well as in Ref.~\cite{OSchutt}) it has been found that the results converge for
the kinetic-energy cutoffs below $60$ Ry.
The cluster is positioned in the center of a large supercell of side $30$ a.u.
Such a choice ensures that the minimum distance between two clusters in the neighboring cells is in the worst case  larger than $9.375$ \AA, which is found to be sufficient to prevent their interaction.
The Brillouin zone is sampled using the
$\Gamma$-point.

Structural relaxation for MgO nanotube clusters is carried out by
performing a series of self-consistent calculations and computing the forces on atoms. Rings and
stable stacks of rings for MgO clusters
have already been suggested by empirical and Hartree-Fock calculations.
\cite{PJZiemann,Moukouri,Noguera,Malliavin,Recio,Puente,Wilson}
Therefore, here
a cylindrical distribution of atoms in the form of stacks of polygons is assumed and interatomic distances are varied to find a minimum energy configuration.
Starting nanotubes are constructed as strictly parallel stacks of regular polygons of the sides of $0.2$ nm, and
at the distance of $0.2$ nm. The experimental nearest-neighbor distance in the bulk magnesium oxide
is $0.21056$ nm. \cite{Wyckoff}
Calculations have been shown that this distance
decreases in nanoparticles. \cite{PJZiemann,Kohler,Moukouri,Noguera,Malliavin,Roberts,Recio,Puente,Wilson,Veliah,FCalvo}
The experimental distance between atoms in the dimer MgO is $0.1749$ nm. \cite{Huber}
Stacks of four parallel squares ($4 \times 4$), hexagons ($6 \times 4$), octagons ($8 \times 4$), and
decagons ($10 \times 4$) are studied. Therefore, the initial lengths of nanotubes are $0.6$ nm.
Several similar nanotubes are made for each size by breaking the symmetry of an ideal starting configuration by small random displacements of atoms. This procedure yields negligible difference in the final energy minima.
The geometry optimizations are performed using the Broyden-Fletcher-Goldfarb-Shanno method of minimization.
All atoms are allowed to relax without any imposed constraint.
The optimizations are carried out till the forces on the atoms are below $10^{-4}$ a.u.
For geometries far from the minimum, the self-consistency cycle continues till the energy differences are less than $10^{-8}$ a.u. Close to the minimum smaller
self-consistency thresholds are taken (down to $10^{-10}$ a.u.)
in order to produce correct forces and small energy differences involved in the relaxation.
\cite{SBaroni}
MgO nanowires and nanotubes synthesized in Refs. \cite{Tang,Yin,Dang,YBLi} are typically several tens of micrometers long, and
their outer diameter is $(15-100)$ nm. These sizes are not accessible for computers used presently.
It is found  that   infinite  periodic MgO    tubes  in   the form   of  stacks
of polygons    similar  to     nanotube    clusters  studied   here,    but  in
the superlattice geometry, are not stable.

\section{Results and discussion}
\label{sec:3}

\begin{figure}
\includegraphics*[scale=.72]{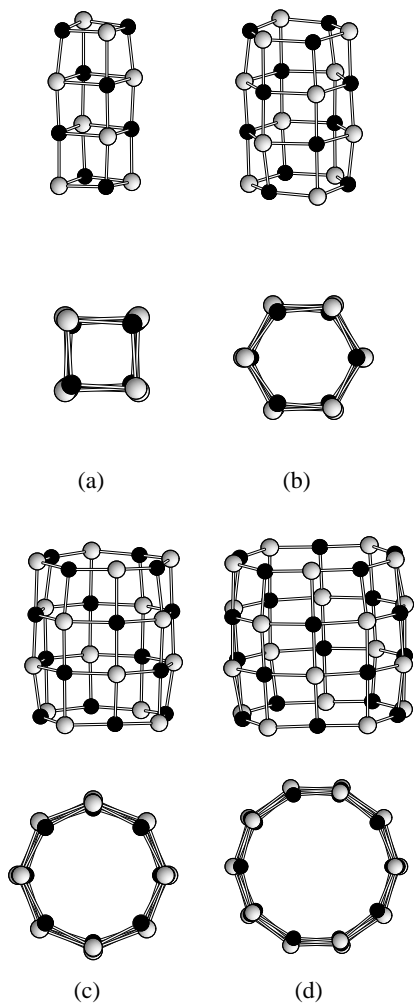}
\caption{The optimized geometries of MgO nanotubes: (a) $4 \times 4$, (b) $6 \times 4$, (c) $8 \times 4$,
(d) $10 \times 4$. Dark and light balls are used to represent the Mg and O ions, respectively.
Side views and views along the tube axis are shown.}
\label{fig:fig1}
\end{figure}

Figure \ref{fig:fig1} presents the structures of MgO nanotubes which correspond to the minimum energy.
This visualization is performed using the Rasmol package.
\cite{Sayle,Bernstein}
The results show that stacked ring structures are stable during a pseudopotential density functional theory
minimization. As presented in Fig.~\ref{fig:fig1},
the bonds and angles slightly change during the geometry optimization. The deviation from planarity of initial
polygons is small.
Similar, but slightly more distorted structures have been obtained by Wilson using an empirical compressible ion potential model.
\cite{Wilson}

It is found here that
the square cross-section of the smallest $4 \times 4$ nanotube evolves to the rhombus. For the
top and bottom rhombuses the O-Mg-O angle is $96.3^{\circ}$, whereas the Mg-O-Mg angle
is $83.5^{\circ}$.
\begin{table}
\caption{\label{tab:table1}The geometry of optimized nanotubes, $a$ is the length of the polygon side, $l$ is the distance
between neighboring polygonal rings, $L$ is the length of the tube (i.e., the distance between opposite atoms on the
top and bottom polygons).
Indices $o$ and $i$ respectively label outer (i.e., when at least one
atom is in the top and bottom polygons) and interior bonds. Units are nm. The range of values is given if
more than two values appear for a certain distance.}
\begin{ruledtabular}
\begin{tabular}{ccccc}
Nanotube&$4 \times 4$ & $6 \times 4$ & $8 \times 4$ & $10 \times 4$\\
\hline
$a_o$ &0.1937          &0.1895; 0.1900  &0.1883; 0.1884 &0.1874-0.1880\\
$a_i$ &0.2101          &0.2029          &0.2007; 0.2009 &0.1997-0.2004\\
$l_o$ &0.1919; 0.1956  &0.1950-0.2000   &0.1961; 0.2017 &0.1963-0.2019\\
$l_i$ &0.1922          &0.1954          &0.1961         &0.1969       \\
$L$     &0.5782          &0.5883; 0.5886  &0.5918         &0.5925       \\
\end{tabular}
\end{ruledtabular}
\end{table}
The $6 \times 4$,  $8 \times 4$, and $10 \times 4$ nanotubes after optimizations have a barrel shape as shown in Fig.~\ref{fig:fig1}.
The optimized structural parameters are shown in Table.~\ref{tab:table1}.
The length of nanotubes increases with the number of atoms in the cluster.
In optimized structures majority of the Mg-O distances decrease in comparison with initial ones of $0.2$ nm.
Total energies  as a function of the initial diameter of MgO cylinders are shown in Table.~\ref{tab:table2}.
\begin{table}
\caption{\label{tab:table2}Total energies per MgO unit. D is the diameter of the enclosed cylinder for an initial regular structure.}
\begin{ruledtabular}
\begin{tabular}{ccc}
Nanotube & Diameter (nm) & Total energies (eV)\\
\hline
$4\times 4$&0.282845 & -464.6981\\
$6 \times 4$&0.400000 & -465.1345\\
$8 \times 4$&0.522603 & -465.2519\\
$10 \times 4$&0.647249 & -465.2947\\
\end{tabular}
\end{ruledtabular}
\end{table}

\begin{figure}
\includegraphics*[scale=.72]{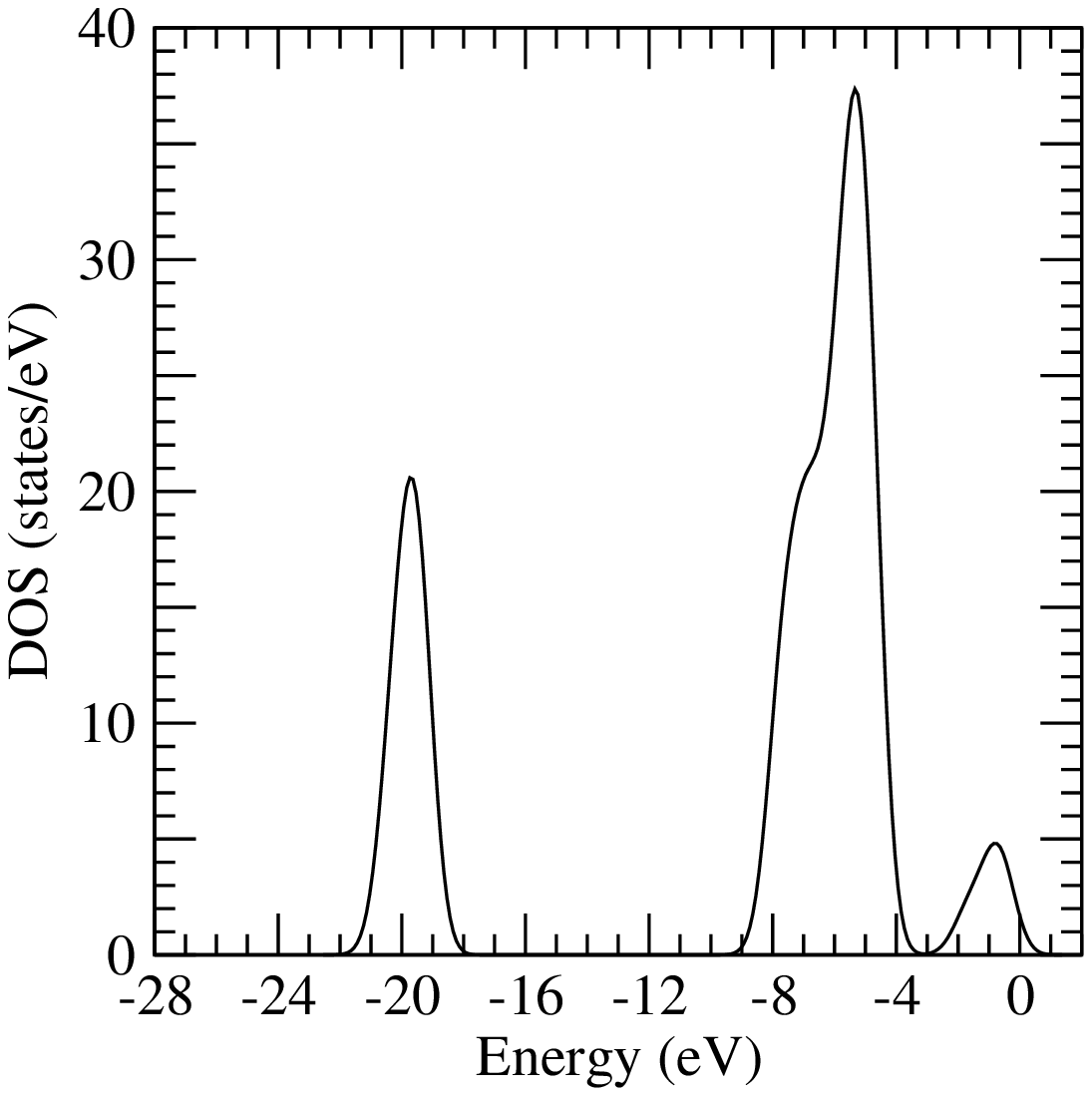}
\caption{Density of states for the $8 \times 4$ MgO nanotube. DOS for other nanotubes is similar.}
\label{fig:fig2}
\end{figure}

The electronic density of states (DOS) for the $8 \times 4$ nanotube is shown in
Fig.~\ref{fig:fig2}.
The separation of bands is about 8.5 eV for all investigated nanotubes.
When the diameter of a nanotube increases DOS does not change substantially, and
the peaks only increase in their heights.
A gap of $5.5$ eV independent of the tube diameter and chirality
has been, for example, found for tubular boron nitride nanotubes.
\cite{Blase}
The band gap value of about $9$ eV has been calculated for the $64$ atoms cubelike MgO cluster using the density functional DMol package.
\cite{Veliah}
However, there the decrease of the band separation when the size of cubes increases has
been reported. It has been calculated that for the (MgO)$_4$ cubelike cluster the band separation is $12$ eV.
The increase of the number of atoms changes the number of
neighbors for a cube cluster geometry where
the coordination in larger clusters approach the bulk value.
Conversely, the coordination in the (MgO)$_n$ nanotube clusters studied here does not approach the bulk value
for large $n$.  A strong shape dependence of electronic properties in small MgO clusters has been found by Moukouri and
Noguera  using a tight binding calculation.
\cite{Moukouri,Noguera}
It is known that both LDA and GGA exchange correlation functionals underestimate the electronic band gap.
The band gap of $7.7$ eV has been calculated
in the GW approximation to the bulk electronic structure of MgO,
in agreement with the experimental value of $7.8$ eV.
\cite{Schonberger}
The LDA approximation used within various methods of electronic structure calculations gives
$(4.19-6.6)$ eV for the bulk band gap of MgO.
\cite{Schonberger}
Therefore, the band gaps of $8.5$ eV calculated in this work for MgO nanotubes within LDA, and  $9$ eV found in
Ref.~\cite{Veliah} for
the cubelike (MgO)$_{32}$ cluster in the same approximation,
are much larger than the values for the LDA bulk gap reported in the literature.
An increase of the band gap in nanoparticles in comparison to the bulk occurs also for other materials.
The origin of this phenomenon is a confinement in finite systems which increases
the separation between energy levels.
The value of the band gap of various MgO clusters has not been measured, and
this subject certainly deserves an investigation.

The L{\" o}wdin population analysis has been carried out.
\cite{Szabo}
The charges on the atoms and
the average charge transfers are shown in Table.~\ref{tab:table3}.
The spilling parameter S measures the difference between the plane-wave eigenstates and their projection into the atomic basis, i.e., the amount of the total charge lost in the projecting process.
\cite{Portal}
Identical atoms with different coordination on the same nanotube sometimes have a different charge.
The large charge transfer, but still less  than
two, shows that  the mixed ionic  and covalent bonding  exist in these  nanotube
clusters. As in Refs. \cite{Moukouri,Noguera,Malliavin},
the  term ionocovalent is used here to  label the mixed ionic and covalent bonding.
\begin{table}
\caption{\label{tab:table3}
The results of the L{\" o}wdin population analysis.
Average charge transfer $\delta Q$ to oxygen atoms is shown for all nanotubes. Q is the charge on the atom whose number
of neighbors is Z (initial valence charge is 2 for Mg, and 6 for O).
S is the spilling parameter.}
\begin{ruledtabular}
\begin{tabular}{ccccc}
Nanotube   &$4 \times 4$ & $6 \times 4$ & $8 \times 4$ & $10 \times 4$\\
\hline
$\delta Q$ &   1.39      & 1.40         &    1.41      &    1.42      \\
Q(Mg, Z=3) &   0.44      & 0.42         &    0.40      &    0.39      \\
Q(Mg, Z=4) &   0.44      & 0.43         &    0.42      &    0.41      \\
Q(O, Z=3)  &   7.37      & 7.39         &    7.40      &    7.41      \\
Q(O, Z=4)  &   7.40      & 7.41         &    7.42      &    7.43      \\
S          &   0.02      & 0.02         &    0.02      &    0.02      \\
\end{tabular}
\end{ruledtabular}
\end{table}
Although in a population analysis and other calculations involving the charge of ions
the values depend upon the calculation model and method, a short discussion of several
results follows. This point strengthens a connection between structural features of nanoparticles
and the charges on the ions.
The analysis of cubelike MgO clusters \cite{Recio} has been
shown that charges depend on the size of nanoparticles and on the coordination number of atoms.
Moukouri and Noguera have been found a large charge reduction as the coordination number decreases yielding
to the charge in the MgO dimer of only half that of the bulk ions.
\cite{Moukouri,Noguera}
In an early study based on the rigid ion approximation and two polarizable ion shell models, Ziemann and Castleman have
found that cubelike structures of (MgO)$_n$ correspond to the model with ion charges $\pm 1$.
\cite{PJZiemann}
In their second model with ion charges $\pm 2$ spherical geometries composed of hexagons and squares have been optimized. Similar results have been obtained by a genetic algorithm method.
\cite{Roberts}
Using an empirical potential with fluctuating charges Calvo has been found that for cubelike (MgO)$_n$ clusters,
when $n$ changes
from small values to the bulk regime, the charge transfer increases from $1$ to $2$.
\cite{FCalvo}
The crossover from ionocovalent to ionic bonds has been estimated to occur in these clusters at  $ (300 \pm 100)$ MgO molecules.
In the nanotube geometry studied here the coordination numbers are the same for all clusters and for larger clusters
do not approach the bulk value.
Therefore, the charge transfer of about $1.4$, and the ionocovalent bonds are typical for these MgO nanotube clusters.
\begin{figure}
\includegraphics*[scale=.72]{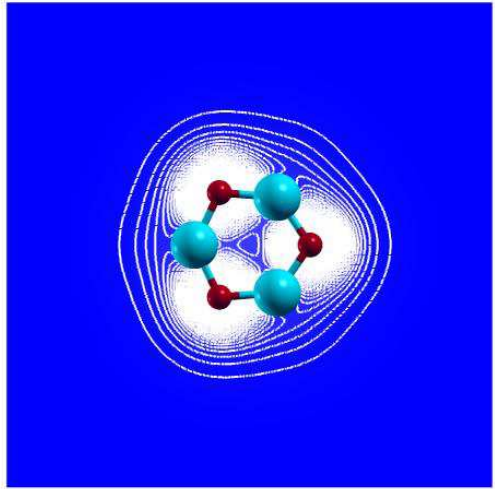}
\caption{The charge density contour plot on a plane perpendicular to the axis of the 6x4 cluster and passing through
the upper hexagon. The $0.01$ $e/a_0^3$ isovalue is shown. The small dark spheres are O, whereas large light ones are Mg atoms.}
\label{fig:fig3}
\end{figure}
The contour  graphical
representation of the charge density for the 6x4 cluster is presented in Fig.~\ref{fig:fig3}.
This visualization  is performed  using the  XCrySDen package. \cite{Tone}
Charges are mostly located on the oxygen atoms,  but the covalent contribution to the  ionic
bonding  exists.

\section{Conclusions}
\label{sec:4}

MgO nanotubes consisting of four stacked planar n-polygonal rings ($n=4,6,8,10$)
are investigated by the first-principles density functional theory
method in the pseudopotential approximation.
These calculations show that MgO nanotube clusters are stable slightly distorted stacks of polygons. They have a large insulating band gap of $8.5$ eV
independently of their radius.
The L{\" o}wdin population analysis indicates that the average charge transfer to oxygen atoms is $1.4$. Therefore, the covalent contribution to the ionic bonding exist in these MgO nanotubes.

Experimental studies of various MgO clusters, for example done by photoemission and electron microscopy
techniques, are desirable.
Recently an atomic resolution imaging of the Mg(001) surface has been reached using a dynamic
scanning force microscopy.
\cite{Barth}
This experiment shows that similar imaging of MgO nanotubes is also feasible.
It will be interesting to prepare, study, and use multiwalled MgO nanowires similar to carbon
\cite{Avouris}
and gold
\cite{Goranka}
ones.
MgO nanotubes are cylindrical structures with large interstitial regions made of a chemically inert material.
Therefore, they are promising for use as nanocapsules.
Submicrometer-sized MgO cubes have already been used as a template for growth of carbon
nanotubes and these structures were imaged by electron microscopy.
\cite{Wei}
It is possible to prepare similar templates based on MgO nanotubes. Rectangular MgO nanorods have been recently used
for the growth of embedded In nanowires.
\cite{Bando}
Such MgO composite nanostructures are promising candidates for antioxidation coating of metal nanowires, and as insulating layers in nanoelectronic devices.
As for the BN nanotubes
\cite{Blase},
a radius independent band gap calculated in this work offers a possibility to use grown samples of
MgO nanotubes containing  different sizes, but with similar electronic properties.
Therefore, many possibilities for synthesis and experimental investigations, as well as for applications
in various fields of nanotechnology exist. Some important applications of MgO  cylindrical nanostructures
have already been described.
\cite{YBLi,PYang,CMLieber}

\begin{acknowledgments}
This work has been carried under the HR-MZT project 0119255 \textquotedblleft Dynamical
Properties and Spectroscopy of Surfaces and Nanostructures\textquotedblright.
The PWscf package has been used. \cite{SBaroni}
Part of the calculations has been done on the
cluster of PCs at the University Computing Center SRCE, Zagreb. I would like to thank Darko Babi\' c
for many advices and discussions about parallel computing. I am also grateful  to  Tone
Kokalj for making available the Cygwin version of his XCrySDen.
\end{acknowledgments}


\begin{thebibliography}{44}
\expandafter\ifx\csname natexlab\endcsname\relax\def\natexlab#1{#1}\fi
\expandafter\ifx\csname bibnamefont\endcsname\relax
  \def\bibnamefont#1{#1}\fi
\expandafter\ifx\csname bibfnamefont\endcsname\relax
  \def\bibfnamefont#1{#1}\fi
\expandafter\ifx\csname citenamefont\endcsname\relax
  \def\citenamefont#1{#1}\fi
\expandafter\ifx\csname url\endcsname\relax
  \def\url#1{\texttt{#1}}\fi
\expandafter\ifx\csname urlprefix\endcsname\relax\def\urlprefix{URL }\fi
\providecommand{\bibinfo}[2]{#2}
\providecommand{\eprint}[2][]{\url{#2}}

\bibitem[{\citenamefont{Dresselhaus et~al.}(2001)\citenamefont{Dresselhaus,
  Dresselhaus, and Avouris}}]{Avouris}
\bibinfo{editor}{\bibfnamefont{M.~S.} \bibnamefont{Dresselhaus}},
  \bibinfo{editor}{\bibfnamefont{G.}~\bibnamefont{Dresselhaus}},
  \bibnamefont{and} \bibinfo{editor}{\bibfnamefont{P.}~\bibnamefont{Avouris}},
  eds., \emph{\bibinfo{title}{Carbon Nanotubes}}
  (\bibinfo{publisher}{Springer}, \bibinfo{address}{Berlin},
  \bibinfo{year}{2001}).

\bibitem[{\citenamefont{Saunders}(1988)}]{Saunders}
\bibinfo{author}{\bibfnamefont{W.~A.} \bibnamefont{Saunders}},
  \bibinfo{journal}{Phys. Rev. B} \textbf{\bibinfo{volume}{37}},
  \bibinfo{pages}{6583} (\bibinfo{year}{1988}).

\bibitem[{\citenamefont{Ziemann and Castleman}(1991)}]{PJZiemann}
\bibinfo{author}{\bibfnamefont{P.~J.} \bibnamefont{Ziemann}} \bibnamefont{and}
  \bibinfo{author}{\bibfnamefont{A.~W.} \bibnamefont{Castleman}},
  \bibinfo{journal}{J. Chem. Phys.} \textbf{\bibinfo{volume}{94}},
  \bibinfo{pages}{718} (\bibinfo{year}{1991}).

\bibitem[{\citenamefont{K{\" o}hler et~al.}(1997)\citenamefont{K{\" o}hler,
  Gail, and Sedlmayr}}]{Kohler}
\bibinfo{author}{\bibfnamefont{T.~M.} \bibnamefont{K{\" o}hler}},
  \bibinfo{author}{\bibfnamefont{H.~P.} \bibnamefont{Gail}}, \bibnamefont{and}
  \bibinfo{author}{\bibfnamefont{E.}~\bibnamefont{Sedlmayr}},
  \bibinfo{journal}{Astron. Astrophys.} \textbf{\bibinfo{volume}{320}},
  \bibinfo{pages}{553} (\bibinfo{year}{1997}).

\bibitem[{\citenamefont{Moukouri and Noguera}(1992)}]{Moukouri}
\bibinfo{author}{\bibfnamefont{S.}~\bibnamefont{Moukouri}} \bibnamefont{and}
  \bibinfo{author}{\bibfnamefont{C.}~\bibnamefont{Noguera}},
  \bibinfo{journal}{Z. Phys. D} \textbf{\bibinfo{volume}{24}},
  \bibinfo{pages}{71} (\bibinfo{year}{1992}).

\bibitem[{\citenamefont{Moukouri and Noguera}(1993)}]{Noguera}
\bibinfo{author}{\bibfnamefont{S.}~\bibnamefont{Moukouri}} \bibnamefont{and}
  \bibinfo{author}{\bibfnamefont{C.}~\bibnamefont{Noguera}},
  \bibinfo{journal}{Z. Phys. D} \textbf{\bibinfo{volume}{27}},
  \bibinfo{pages}{79} (\bibinfo{year}{1993}).

\bibitem[{\citenamefont{Malliavin and Coudray}(1997)}]{Malliavin}
\bibinfo{author}{\bibfnamefont{M.~J.} \bibnamefont{Malliavin}}
  \bibnamefont{and} \bibinfo{author}{\bibfnamefont{C.}~\bibnamefont{Coudray}},
  \bibinfo{journal}{J. Chem. Phys.} \textbf{\bibinfo{volume}{106}},
  \bibinfo{pages}{2323} (\bibinfo{year}{1997}).

\bibitem[{\citenamefont{Roberts and Johnston}(2001)}]{Roberts}
\bibinfo{author}{\bibfnamefont{C.}~\bibnamefont{Roberts}} \bibnamefont{and}
  \bibinfo{author}{\bibfnamefont{R.~L.} \bibnamefont{Johnston}},
  \bibinfo{journal}{Phys. Chem. Chem. Phys.} \textbf{\bibinfo{volume}{3}},
  \bibinfo{pages}{5024} (\bibinfo{year}{2001}).

\bibitem[{\citenamefont{Recio et~al.}(1993)\citenamefont{Recio, Pandey, Ayuela,
  and Kunz}}]{Recio}
\bibinfo{author}{\bibfnamefont{J.~M.} \bibnamefont{Recio}},
  \bibinfo{author}{\bibfnamefont{R.}~\bibnamefont{Pandey}},
  \bibinfo{author}{\bibfnamefont{A.}~\bibnamefont{Ayuela}}, \bibnamefont{and}
  \bibinfo{author}{\bibfnamefont{A.~B.} \bibnamefont{Kunz}},
  \bibinfo{journal}{J. Chem. Phys.} \textbf{\bibinfo{volume}{98}},
  \bibinfo{pages}{4783} (\bibinfo{year}{1993}).

\bibitem[{\citenamefont{de~la Puente et~al.}(1997)\citenamefont{de~la Puente,
  Aguado, Ayuela, and Lopez}}]{Puente}
\bibinfo{author}{\bibfnamefont{E.}~\bibnamefont{de~la Puente}},
  \bibinfo{author}{\bibfnamefont{A.}~\bibnamefont{Aguado}},
  \bibinfo{author}{\bibfnamefont{A.}~\bibnamefont{Ayuela}}, \bibnamefont{and}
  \bibinfo{author}{\bibfnamefont{J.~M.} \bibnamefont{Lopez}},
  \bibinfo{journal}{Phys. Rev. B} \textbf{\bibinfo{volume}{56}},
  \bibinfo{pages}{7607} (\bibinfo{year}{1997}).

\bibitem[{\citenamefont{Wilson}(1997)}]{Wilson}
\bibinfo{author}{\bibfnamefont{M.}~\bibnamefont{Wilson}}, \bibinfo{journal}{J.
  Chem. Phys.} \textbf{\bibinfo{volume}{101}}, \bibinfo{pages}{4917}
  (\bibinfo{year}{1997}).

\bibitem[{\citenamefont{Veliah et~al.}(1995)\citenamefont{Veliah, Pandey, Li,
  Newsam, and Vessal}}]{Veliah}
\bibinfo{author}{\bibfnamefont{S.}~\bibnamefont{Veliah}},
  \bibinfo{author}{\bibfnamefont{R.}~\bibnamefont{Pandey}},
  \bibinfo{author}{\bibfnamefont{Y.~S.} \bibnamefont{Li}},
  \bibinfo{author}{\bibfnamefont{J.~M.} \bibnamefont{Newsam}},
  \bibnamefont{and} \bibinfo{author}{\bibfnamefont{B.}~\bibnamefont{Vessal}},
  \bibinfo{journal}{Chem. Phys. Lett.} \textbf{\bibinfo{volume}{235}},
  \bibinfo{pages}{53} (\bibinfo{year}{1995}).

\bibitem[{\citenamefont{Calvo}(2003)}]{FCalvo}
\bibinfo{author}{\bibfnamefont{F.}~\bibnamefont{Calvo}},
  \bibinfo{journal}{Phys. Rev. B} \textbf{\bibinfo{volume}{67}},
  \bibinfo{pages}{161403} (\bibinfo{year}{2003}).

\bibitem[{\citenamefont{Tang et~al.}(2002)\citenamefont{Tang, Bando, and
  Sato}}]{Tang}
\bibinfo{author}{\bibfnamefont{C.}~\bibnamefont{Tang}},
  \bibinfo{author}{\bibfnamefont{Y.}~\bibnamefont{Bando}}, \bibnamefont{and}
  \bibinfo{author}{\bibfnamefont{T.}~\bibnamefont{Sato}}, \bibinfo{journal}{J.
  Phys. Chem B} \textbf{\bibinfo{volume}{106}}, \bibinfo{pages}{7449}
  (\bibinfo{year}{2002}).

\bibitem[{\citenamefont{Yin et~al.}(2002)\citenamefont{Yin, Zhang, and
  Xia}}]{Yin}
\bibinfo{author}{\bibfnamefont{Y.}~\bibnamefont{Yin}},
  \bibinfo{author}{\bibfnamefont{G.}~\bibnamefont{Zhang}}, \bibnamefont{and}
  \bibinfo{author}{\bibfnamefont{Y.}~\bibnamefont{Xia}}, \bibinfo{journal}{Adv.
  Funct. Mater.} \textbf{\bibinfo{volume}{12}}, \bibinfo{pages}{293}
  (\bibinfo{year}{2002}).

\bibitem[{\citenamefont{Dang et~al.}(2003)\citenamefont{Dang, Wang, and
  Fan}}]{Dang}
\bibinfo{author}{\bibfnamefont{H.~Y.} \bibnamefont{Dang}},
  \bibinfo{author}{\bibfnamefont{J.}~\bibnamefont{Wang}}, \bibnamefont{and}
  \bibinfo{author}{\bibfnamefont{S.~S.} \bibnamefont{Fan}},
  \bibinfo{journal}{Nanotechnology} \textbf{\bibinfo{volume}{14}},
  \bibinfo{pages}{738} (\bibinfo{year}{2003}).

\bibitem[{\citenamefont{Li et~al.}(2003)\citenamefont{Li, Bando, Golberg, and
  Liu}}]{YBLi}
\bibinfo{author}{\bibfnamefont{Y.~B.} \bibnamefont{Li}},
  \bibinfo{author}{\bibfnamefont{Y.}~\bibnamefont{Bando}},
  \bibinfo{author}{\bibfnamefont{D.}~\bibnamefont{Golberg}}, \bibnamefont{and}
  \bibinfo{author}{\bibfnamefont{Z.~W.} \bibnamefont{Liu}},
  \bibinfo{journal}{Appl. Phys. Lett.} \textbf{\bibinfo{volume}{83}},
  \bibinfo{pages}{999} (\bibinfo{year}{2003}).

\bibitem[{\citenamefont{Gao and Bando}(2002)}]{YGao}
\bibinfo{author}{\bibfnamefont{Y.}~\bibnamefont{Gao}} \bibnamefont{and}
  \bibinfo{author}{\bibfnamefont{Y.}~\bibnamefont{Bando}},
  \bibinfo{journal}{Nature} \textbf{\bibinfo{volume}{415}},
  \bibinfo{pages}{599} (\bibinfo{year}{2002}).

\bibitem[{\citenamefont{Yang and Lieber}(1996)}]{PYang}
\bibinfo{author}{\bibfnamefont{P.}~\bibnamefont{Yang}} \bibnamefont{and}
  \bibinfo{author}{\bibfnamefont{C.~M.} \bibnamefont{Lieber}},
  \bibinfo{journal}{Science} \textbf{\bibinfo{volume}{273}},
  \bibinfo{pages}{1836} (\bibinfo{year}{1996}).

\bibitem[{\citenamefont{Yang and Lieber}(1997)}]{CMLieber}
\bibinfo{author}{\bibfnamefont{P.}~\bibnamefont{Yang}} \bibnamefont{and}
  \bibinfo{author}{\bibfnamefont{C.~M.} \bibnamefont{Lieber}},
  \bibinfo{journal}{J. Mater. Res.} \textbf{\bibinfo{volume}{12}},
  \bibinfo{pages}{2981} (\bibinfo{year}{1997}).

\bibitem[{\citenamefont{Hohenberg and Kohn}(1964)}]{Kohn}
\bibinfo{author}{\bibfnamefont{P.}~\bibnamefont{Hohenberg}} \bibnamefont{and}
  \bibinfo{author}{\bibfnamefont{W.}~\bibnamefont{Kohn}},
  \bibinfo{journal}{Phys. Rev.} \textbf{\bibinfo{volume}{136}},
  \bibinfo{pages}{B864} (\bibinfo{year}{1964}).

\bibitem[{\citenamefont{Kohn and Sham}(1965)}]{Sham}
\bibinfo{author}{\bibfnamefont{W.}~\bibnamefont{Kohn}} \bibnamefont{and}
  \bibinfo{author}{\bibfnamefont{L.~J.} \bibnamefont{Sham}},
  \bibinfo{journal}{Phys. Rev.} \textbf{\bibinfo{volume}{140}},
  \bibinfo{pages}{A1133} (\bibinfo{year}{1965}).

\bibitem[{\citenamefont{Baroni et~al.}(2001)\citenamefont{Baroni, de~Gironcoli,
  dal Corso, and Giannozzi}}]{SdeGironcoli}
\bibinfo{author}{\bibfnamefont{S.}~\bibnamefont{Baroni}},
  \bibinfo{author}{\bibfnamefont{S.}~\bibnamefont{de~Gironcoli}},
  \bibinfo{author}{\bibfnamefont{A.}~\bibnamefont{dal Corso}},
  \bibnamefont{and}
  \bibinfo{author}{\bibfnamefont{P.}~\bibnamefont{Giannozzi}},
  \bibinfo{journal}{Rev. Mod. Phys.} \textbf{\bibinfo{volume}{73}},
  \bibinfo{pages}{515} (\bibinfo{year}{2001}).

\bibitem[{\citenamefont{Baroni et~al.}()\citenamefont{Baroni, dal Corso,
  de~Gironcoli, and Giannozzi}}]{SBaroni}
\bibinfo{author}{\bibfnamefont{S.}~\bibnamefont{Baroni}},
  \bibinfo{author}{\bibfnamefont{A.}~\bibnamefont{dal Corso}},
  \bibinfo{author}{\bibfnamefont{S.}~\bibnamefont{de~Gironcoli}},
  \bibnamefont{and}
  \bibinfo{author}{\bibfnamefont{P.}~\bibnamefont{Giannozzi}},
  \urlprefix\url{http://www.pwscf.org}.

\bibitem[{\citenamefont{Karki et~al.}(1999)\citenamefont{Karki, Wentzcovitch,
  de~Gironcoli, and Baroni}}]{BBKarki}
\bibinfo{author}{\bibfnamefont{B.~B.} \bibnamefont{Karki}},
  \bibinfo{author}{\bibfnamefont{R.~M.} \bibnamefont{Wentzcovitch}},
  \bibinfo{author}{\bibfnamefont{S.}~\bibnamefont{de~Gironcoli}},
  \bibnamefont{and} \bibinfo{author}{\bibfnamefont{S.}~\bibnamefont{Baroni}},
  \bibinfo{journal}{Science} \textbf{\bibinfo{volume}{286}},
  \bibinfo{pages}{1705} (\bibinfo{year}{1999}).

\bibitem[{\citenamefont{Karki et~al.}(2000)\citenamefont{Karki, Wentzcovitch,
  de~Gironcoli, and Baroni}}]{Wentzcovitch}
\bibinfo{author}{\bibfnamefont{B.~B.} \bibnamefont{Karki}},
  \bibinfo{author}{\bibfnamefont{R.~M.} \bibnamefont{Wentzcovitch}},
  \bibinfo{author}{\bibfnamefont{S.}~\bibnamefont{de~Gironcoli}},
  \bibnamefont{and} \bibinfo{author}{\bibfnamefont{S.}~\bibnamefont{Baroni}},
  \bibinfo{journal}{Phys. Rev. B} \textbf{\bibinfo{volume}{61}},
  \bibinfo{pages}{8793} (\bibinfo{year}{2000}).

\bibitem[{\citenamefont{Oganov and Dorogokupets}(2003)}]{Oganov}
\bibinfo{author}{\bibfnamefont{A.~R.} \bibnamefont{Oganov}} \bibnamefont{and}
  \bibinfo{author}{\bibfnamefont{P.~I.} \bibnamefont{Dorogokupets}},
  \bibinfo{journal}{Phys. Rev. B} \textbf{\bibinfo{volume}{67}},
  \bibinfo{pages}{224110} (\bibinfo{year}{2003}).

\bibitem[{\citenamefont{Sch{\" u}tt et~al.}(1994)\citenamefont{Sch{\" u}tt,
  Pavone, Windl, Karch, and Strauch}}]{OSchutt}
\bibinfo{author}{\bibfnamefont{O.}~\bibnamefont{Sch{\" u}tt}},
  \bibinfo{author}{\bibfnamefont{P.}~\bibnamefont{Pavone}},
  \bibinfo{author}{\bibfnamefont{W.}~\bibnamefont{Windl}},
  \bibinfo{author}{\bibfnamefont{K.}~\bibnamefont{Karch}}, \bibnamefont{and}
  \bibinfo{author}{\bibfnamefont{D.}~\bibnamefont{Strauch}},
  \bibinfo{journal}{Phys. Rev. B} \textbf{\bibinfo{volume}{50}},
  \bibinfo{pages}{3746} (\bibinfo{year}{1994}).

\bibitem[{\citenamefont{Perdew and Zunger}(1981)}]{Perdew}
\bibinfo{author}{\bibfnamefont{J.~P.} \bibnamefont{Perdew}} \bibnamefont{and}
  \bibinfo{author}{\bibfnamefont{A.}~\bibnamefont{Zunger}},
  \bibinfo{journal}{Phys. Rev. B} \textbf{\bibinfo{volume}{23}},
  \bibinfo{pages}{5048} (\bibinfo{year}{1981}).

\bibitem[{\citenamefont{von Barth and Car}()}]{Car}
\bibinfo{author}{\bibfnamefont{U.}~\bibnamefont{von Barth}} \bibnamefont{and}
  \bibinfo{author}{\bibfnamefont{R.}~\bibnamefont{Car}},
  \bibinfo{note}{unpublished}.

\bibitem[{\citenamefont{Troullier and Martins}(1991)}]{Martins}
\bibinfo{author}{\bibfnamefont{N.}~\bibnamefont{Troullier}} \bibnamefont{and}
  \bibinfo{author}{\bibfnamefont{J.~L.} \bibnamefont{Martins}},
  \bibinfo{journal}{Phys. Rev. B} \textbf{\bibinfo{volume}{43}},
  \bibinfo{pages}{1993} (\bibinfo{year}{1991}).

\bibitem[{\citenamefont{Wyckoff}(1963)}]{Wyckoff}
\bibinfo{author}{\bibfnamefont{R.~W.~G.} \bibnamefont{Wyckoff}},
  \emph{\bibinfo{title}{Crystal Structures}} (\bibinfo{publisher}{Wiley},
  \bibinfo{address}{New York}, \bibinfo{year}{1963}).

\bibitem[{\citenamefont{Huber and Herzberg}(1979)}]{Huber}
\bibinfo{author}{\bibfnamefont{K.~P.} \bibnamefont{Huber}} \bibnamefont{and}
  \bibinfo{author}{\bibfnamefont{G.}~\bibnamefont{Herzberg}},
  \emph{\bibinfo{title}{Molecular Spectra and Molecular Structure: IV Constants
  of Diatomic Molecules}} (\bibinfo{publisher}{Van Nostrand Reinhold},
  \bibinfo{address}{New York}, \bibinfo{year}{1979}).

\bibitem[{\citenamefont{Sayle and Milner-White}(1995)}]{Sayle}
\bibinfo{author}{\bibfnamefont{R.}~\bibnamefont{Sayle}} \bibnamefont{and}
  \bibinfo{author}{\bibfnamefont{E.~J.} \bibnamefont{Milner-White}},
  \bibinfo{journal}{Trends Biochem. Sci.} \textbf{\bibinfo{volume}{20}},
  \bibinfo{pages}{374} (\bibinfo{year}{1995}).

\bibitem[{\citenamefont{Bernstein}(2000)}]{Bernstein}
\bibinfo{author}{\bibfnamefont{H.~J.} \bibnamefont{Bernstein}},
  \bibinfo{journal}{Trends Biochem. Sci.} \textbf{\bibinfo{volume}{25}},
  \bibinfo{pages}{453} (\bibinfo{year}{2000}).

\bibitem[{\citenamefont{Blase et~al.}(1994)\citenamefont{Blase, Rubio, Louie,
  and Cohen}}]{Blase}
\bibinfo{author}{\bibfnamefont{X.}~\bibnamefont{Blase}},
  \bibinfo{author}{\bibfnamefont{A.}~\bibnamefont{Rubio}},
  \bibinfo{author}{\bibfnamefont{S.~G.} \bibnamefont{Louie}}, \bibnamefont{and}
  \bibinfo{author}{\bibfnamefont{M.~L.} \bibnamefont{Cohen}},
  \bibinfo{journal}{Europhys. Lett.} \textbf{\bibinfo{volume}{28}},
  \bibinfo{pages}{335} (\bibinfo{year}{1994}).

\bibitem[{\citenamefont{Sch{\" o}nberger and Aryasetiawan}(1995)}]{Schonberger}
\bibinfo{author}{\bibfnamefont{U.}~\bibnamefont{Sch{\" o}nberger}}
  \bibnamefont{and}
  \bibinfo{author}{\bibfnamefont{F.}~\bibnamefont{Aryasetiawan}},
  \bibinfo{journal}{Phys. Rev. B} \textbf{\bibinfo{volume}{52}},
  \bibinfo{pages}{8788} (\bibinfo{year}{1995}).

\bibitem[{\citenamefont{Szabo and Ostlund}(1996)}]{Szabo}
\bibinfo{author}{\bibfnamefont{A.}~\bibnamefont{Szabo}} \bibnamefont{and}
  \bibinfo{author}{\bibfnamefont{N.~S.} \bibnamefont{Ostlund}},
  \emph{\bibinfo{title}{Modern Quantum Chemistry}} (\bibinfo{publisher}{Dover},
  \bibinfo{address}{New York}, \bibinfo{year}{1996}).

\bibitem[{\citenamefont{Sanchez-Portal
  et~al.}(1995)\citenamefont{Sanchez-Portal, Artacho, and Soler}}]{Portal}
\bibinfo{author}{\bibfnamefont{D.}~\bibnamefont{Sanchez-Portal}},
  \bibinfo{author}{\bibfnamefont{E.}~\bibnamefont{Artacho}}, \bibnamefont{and}
  \bibinfo{author}{\bibfnamefont{J.~M.} \bibnamefont{Soler}},
  \bibinfo{journal}{Solid St. Commun.} \textbf{\bibinfo{volume}{95}},
  \bibinfo{pages}{685} (\bibinfo{year}{1995}).

\bibitem[{\citenamefont{Tone}(1999)}]{Tone}
\bibinfo{author}{\bibfnamefont{A.}~\bibnamefont{Kokalj}},
  \bibinfo{journal}{J. Mol. Graphics Modelling} \textbf{\bibinfo{volume}{17}},
  \bibinfo{pages}{176} (\bibinfo{year}{1999}),
  \urlprefix\url{http://www.xcrysden.org}.

\bibitem[{\citenamefont{Barth and Henry}(2003)}]{Barth}
\bibinfo{author}{\bibfnamefont{C.}~\bibnamefont{Barth}} \bibnamefont{and}
  \bibinfo{author}{\bibfnamefont{C.~R.} \bibnamefont{Henry}},
  \bibinfo{journal}{Phys. Rev. Lett} \textbf{\bibinfo{volume}{91}},
  \bibinfo{pages}{196102} (\bibinfo{year}{2003}).

\bibitem[{\citenamefont{Bilalbegovi{\' c}}(1998)}]{Goranka}
\bibinfo{author}{\bibfnamefont{G.}~\bibnamefont{Bilalbegovi{\' c}}},
  \bibinfo{journal}{Phys. Rev. B} \textbf{\bibinfo{volume}{58}},
  \bibinfo{pages}{15412} (\bibinfo{year}{1998}).

\bibitem[{\citenamefont{Wei et~al.}(2001)\citenamefont{Wei, Vajtai, Zhang,
  Ramanath, and Ajayan}}]{Wei}
\bibinfo{author}{\bibfnamefont{B.~Q.} \bibnamefont{Wei}},
  \bibinfo{author}{\bibfnamefont{R.}~\bibnamefont{Vajtai}},
  \bibinfo{author}{\bibfnamefont{Z.~J.} \bibnamefont{Zhang}},
  \bibinfo{author}{\bibfnamefont{G.}~\bibnamefont{Ramanath}}, \bibnamefont{and}
  \bibinfo{author}{\bibfnamefont{P.~M.} \bibnamefont{Ajayan}},
  \bibinfo{journal}{J. Nanosci. Nanotech.} \textbf{\bibinfo{volume}{1}},
  \bibinfo{pages}{35} (\bibinfo{year}{2001}).

\bibitem[{\citenamefont{Tang et~al.}(2003)\citenamefont{Tang, Bourgeois, Bando,
  and Golberg}}]{Bando}
\bibinfo{author}{\bibfnamefont{C.}~\bibnamefont{Tang}},
  \bibinfo{author}{\bibfnamefont{L.}~\bibnamefont{Bourgeois}},
  \bibinfo{author}{\bibfnamefont{Y.}~\bibnamefont{Bando}}, \bibnamefont{and}
  \bibinfo{author}{\bibfnamefont{D.}~\bibnamefont{Golberg}},
  \bibinfo{journal}{Chem. Phys. Lett.} \textbf{\bibinfo{volume}{382}},
  \bibinfo{pages}{374} (\bibinfo{year}{2003}).

\end{thebibliography}
\end{document}